\documentclass[aps,prl,12pt,showpacs,preprintnumbers,nofootinbib,superscriptaddress]{revtex4}
\usepackage{lscape}
\usepackage{indentfirst}
\usepackage{latexsym}
\usepackage{multirow}

\usepackage{amssymb,amsmath}
\usepackage{graphicx}
\newcommand{\bea}{\begin{eqnarray}}
\newcommand{\eea}{\end{eqnarray}}
\newcommand{\be}{\begin{equation}}
\newcommand{\ee}{\end{equation}}
\newcommand{\bes}{\begin{subequations}}
\newcommand{\ees}{\end{subequations}}
\def\lag{\langle}
\def\rag{\rangle}

\def\nn{\nonumber \\}

\begin{document}

\title{Non-Gaussian Stochastic Gravity}

\author{Jason D. Bates}
\email{jdbates@staff.tku.edu.tw}
\affiliation{Department of Physics, Tamkang University, Tamsui, New Taipei City, Taiwan}

\date{May 16, 2013}

\begin{abstract}

This paper presents a new, non-Gaussian formulation of stochastic gravity by incorporating the higher moments of the fluctuations of the quantum stress energy tensor for a free quantum scalar field in a consistent way.  A scheme is developed for obtaining realizations of these fluctuations in terms of the Wightman function, and the behavior of the fluctuations is investigated. The resulting probability distribution for fluctuations of the energy density in Minkowski spacetime is found to be similar to a shifted Gamma distribution.  This distribution features a minimum energy density cutoff at a small negative value, but a sharp peak in the vicinity of this cutoff such that the total probability of observing a negative value is approximately 62\%, balanced by correspondingly larger but rarer positive values.

\end{abstract}


\maketitle

Stochastic gravity~\cite{stograLivRev} is a theoretical framework in which the backreaction problem of semiclassical gravity is treated as a quantum open system, with quantum matter fields playing the role of the environment and the metric field that of the system of interest.  The motivation for this treatment is the similarity between the closed time path (CTP) formalism~\cite{CTP} for computing in-in expectation values within a background spacetime and the influence functional approach of Feynman and Vernon~\cite{FV63}.  As discussed in~\cite{calhu94}, this similarity provides insight into the physical meaning of the noise and dissipation kernels present in the CTP effective action.

The key insight is the identification of the noise kernel, contained in the imaginary portion of the effective action, with the characteristic function of a classical stochastic tensor field which contains all of the information regarding the quantum fluctuations of the matter fields.  By writing an "improved" effective action in terms of this stochastic tensor field and varying with respect to the metric one may derive~\cite{martin99b} a semiclassical Langevin-like equation for the metric perturbations induced by the quantum matter fields.

One of the difficulties encountered in this approach is finding solutions for the influence functional.  For some simplified models in which the environment is linearly coupled to the system of interest, such as those looked at in~\cite{CL83}, exact solutions are possible.  In those cases, the effective action contains only terms of quadratic order, and the imaginary part is exactly what one would expect from the characteristic function of a Gaussian stochastic process.  However, for more general models (especially those with non-linear coupling between the system and the environment, as is the case with gravity) the effective action may contain terms beyond quadratic order which would lead to non-Gaussianities of the stochastic field.  To the author's knowledge, no systematic attempt to treat these higher order terms has been made;~\cite{martin99b} dealt with the issue by simply truncating a functional Taylor series expansion of the influence functional at quadratic order and assuming Gaussianity on the part of the stress tensor fluctuations.  While this may be sufficient to shed light on the nature of the fluctuations as a first approximation, more recent work~\cite{FFR10,FFR12} indicates that the proper probability distribution for fluctuations of the stress energy tensor is, in fact, highly non-Gaussian.

The purpose of this paper is to present a new formulation of stochastic gravity which incorporates this non-Gaussian behavior.  However, if we relax the assumption of Gaussianity, we find that there is an infinity of possible theories of stochastic gravity based upon the choice of the higher moments of the fluctuations.  While the correct choice would depend upon finding an exact expression for the influence functional, in the absence of such an expression we conjecture that the proper choice of the moments is determined by the n-point correlation functions of the stress tensor operator.  There is some evidence which suggests that this conjecture is the correct one, which we present below.  Next, we construct a new stochastic field which satisfies our conjecture and show a method for finding coarse-grained realizations of this field.  Finally, we apply this method to the case of a massless quantum scalar field living in Minkowski spacetime and derive expressions for the probability density function of the field squared and the stress tensor.  Although coarse-graining is through the use of smearing functions is necessary to regulate the n-point functions of the stress tensor, the probability density functions (when expressed in terms of dimensionless quantities) are independent of the choice of smearing.

We begin by considering the CTP action for a metric field $g_{ab}$ interacting with a scalar field $\phi(x)$, which is given by
\bea
  e^{i S_{CTP}[g^+,g^-]} &=& \int D\phi^+ D\phi^- \rho(\phi^+, \phi^-) e^{i \{S_g[g^+] - S_g[g^-]+ S_m[\phi^+,g^+]-S_m[\phi^-,g^-]\}} \nn
                                            &=& e^{i \{S_g[g^+] - S_g[g^-]\}}\mathcal{F}_{IF}[g^+,g^-] \;, \label{CTP}
\eea
where $S_g$ is the standard Einstein-Hilbert action and $S_m$ is the action of the scalar field.  Noting that the stress tensor operator may be obtained by variation of $S_m$ with respect to the metric,
\bea
  \frac{2}{\sqrt{-g(x)}}\frac{\delta S_m}{\delta g_{ab}(x)} &=& \hat{T}_{ab}(x) \; ,
\eea
and writing the metric in terms of perturbation on a background spacetime, $g_{ab} \rightarrow g_{ab} + h_{ab}$, this action may be expanded perturbatively as~\cite{martin99b}
\bea
S_{CTP}[g+h^+,g+h^-] &=& S_g[g+h^+] - S_g[g+h^-] + \frac{1}{2} \int d^4x \, \sqrt{-g(x)} \, \Delta h^{ab}(x) \lag \hat{T}_{ab}(x) \rag \nn
                                        & & \!\!\!\!\!\!\!\!\!\!\!\! + \frac{1}{2} \int d^4x \, d^4x' \, \sqrt{-g(x)}\sqrt{-g(x')} \Sigma h^{ab}(x) D_{abc'd'}(x,x') \Delta h^{c'd'}(x') \nn
                                        & & \!\!\!\!\!\!\!\!\!\!\!\! + \frac{i}{2} \int d^4x \, d^4x' \, \sqrt{-g(x)}\sqrt{-g(x')} \Delta h^{ab}(x) N_{abc'd'}(x,x') \Delta h^{c'd'}(x') \nn
                                        & & \!\!\!\!\!\!\!\!\!\!\!\! + \,  O(h^3) \;. \label{CTP-eff}
\eea
Here $S_g$ is the standard gravitational action, $\Delta h = h^+ - h^-$, and $\Sigma h = h^+ + h^-$. $N_{abc'd'}$ and $D_{abc'd'}$ are the noise and dissipation kernels, respectively.  Explicitly, the noise kernel $N_{abc'd'}$ is given by the symmetrized two point function of the quantum stress energy fluctuation operator,
\bea
  N_{abc'd'}(x,x') = \lag \{ \hat{t}_{ab}(x), \hat{t}_{c'd'}(x') \} \rag \;, \label{nk-def}
\eea
with $\hat{t}_{ab} = \hat{T}_{ab} - \lag \hat{T}_{ab} \rag$.  The dissipation kernel may be obtained~\cite{martin99b} from the noise kernel via the fluctuation-dissipation relation.

For the Gaussian theory described in~\cite{martin99b}, Eq.~\eqref{CTP-eff} is truncated at quadratic order and the noise kernel is identified with the characteristic function of a Gaussian stochastic tensor field $\xi_{ab}$,
\bea
  \lag e^{i \int d^4x \sqrt{-g(x)} \Delta h^{ab}(x) \xi_{ab}(x)} \rag_s &\equiv& e^{ -\frac{1}{2} \int d^4x \, d^4x' \, \sqrt{-g(x)}\sqrt{-g(x')} \Delta h^{ab}(x) N_{abc'd'}(x,x') \Delta h^{c'd'}(x')} \;, \label{xi-gauss}
\eea
where $\lag...\rag_s$ is used to indicate a stochastic, rather than quantum, expectation value.
This allows us to write an ``improved'' effective action,
\bea
S_{eff}[g+h^+,g+h^-] &=& S_g[g+h^+] - S_g[g+h^-] + \frac{1}{2} \int d^4x \, \sqrt{-g(x)} \, \Delta h^{ab}(x) [\lag \hat{T}_{ab}(x) \rag + \xi_{ab}] \nn
                                        & & \!\!\!\!\!\!\!\!\!\!\!\! + \frac{1}{2} \int d^4x \, d^4x' \, \sqrt{-g(x)}\sqrt{-g(x')} \Sigma h^{ab}(x) D_{abc'd'}(x,x') \Delta h^{c'd'}(x') \;,
\eea
where $\xi_{ab}$ is uniquely defined by its first two moments,
\bes
\bea
  \lag \xi_{ab}(x) \rag_s &\equiv& 0 \\
  \lag \xi_{ab}(x) \xi_{c'd'}(x') \rag_s &\equiv& N_{abc'd'}(x,x')  \;.
\eea \label{xi_def}
\ees
Varying this action with respect to the metric gives the Einstein-Langevin equation~\cite{ELE} of stochastic gravity,
\bea 
  G_{ab}(x) = 8\pi [\langle \hat{T}_{ab}(x)\rangle+\xi_{ab}(x)] \;, \label{e-l}
\eea
where the role of the stochastic source term $\xi_{ab}$ as the generator of metric fluctuations is clear.

However, if we consider non-Gaussian $\xi_{ab}$, Eq.~\eqref{xi-gauss} no longer holds, and the characteristic function should contain terms beyond quadratic order in $\Delta h$.  Rather, the higher moments of $\xi_{ab}$ would be determined by the higher order terms in the expansion of the influence functional.  Considering only those terms proportional to some power of $\Delta h$, each order of the expansion of Eq.~\eqref{CTP-eff} may be written as\footnote{This may be seen by noting that the real part of the n-point functions of $\hat{t}_{ab}$ is symmetric under interchange of points, whereas the imaginary part is anti-symmetric.  As only the symmetric part contributes to the terms proportional to $\Delta h^n$, the time ordering usually present in the path integral may be neglected for those terms.}
\bea
  \frac{i^n}{n!}\int d^4x_1 ... d^4x_n \sqrt{-g(x_1)}...\sqrt{-g(x_n)} \Delta h^{a_1b_1}(x_1)...\Delta h^{a_nb_n}(x_n) \lag \{ \hat{t}_{a_1b_1}(x_1),...,\hat{t}_{a_nb_n}(x_n) \} \rag \;, \label {CTP-n}
\eea
where the braces indicate symmetrization of points.  Thus, again considering only those terms proportional to powers of $\Delta h$, each such term may be accounted for by redefining $\xi_{ab}$ using the conjecture
\bea
  \lag e^{i \int d^4x \sqrt{-g(x)} \Delta h^{ab}(x) \xi_{ab}(x)} \rag_s &\equiv& \lag e^{i \int d^4x \sqrt{-g(x)} \Delta h^{ab}(x)\hat{t}_{ab}(x)} \rag \;. \label{xi-ng}
\eea
This gives some evidence that the conjecture used here is the correct one; however, there will, of course, be other terms present at each order in the expansion of Eq.~\eqref{CTP-eff} proportional to some combination of powers of $\Delta h$ and $\Sigma h$.  While it is the suspicion of the author that each of those terms may be related to Eq.~\eqref{CTP-n} by some form of a fluctuation-dissipation relation, the exact meaning is unclear.

The next task is to find realizations of $\xi_{ab}$ based upon the definition given in Eq.~\eqref{xi-ng}.  The procedure used here will parallel that of~\cite{Bates13} for Gaussian fluctuations.  However, as it is difficult to make use of Eq.~\eqref{xi-ng} directly, we begin by defining a new Gaussian stochastic scalar field $\zeta(x)$ using
\bes
\bea
  \lag \zeta(x) \rag_s &\equiv& 0 \\
  \lag \zeta(x) \zeta(x') \rag_s &\equiv& \lag \hat{\phi}(x)\hat{\phi}(x') \rag \;.
\eea \label{xi_def}
\ees
For free scalar fields, Wick's theorem tells us that the n-point correlation functions of $\hat{\phi}$ are exactly those one would expect from a Gaussian stochastic process; thus, the moments of $\zeta$ recreate all of the n-point correlators of the field,
\bea
  \lag \zeta(x_1)...\zeta(x_n) \rag_{s} &=& \lag \{ \hat{\phi}(x_1),..., \hat{\phi}(x_n)\} \rag \;. \label{zeta_moments}
\eea

Using the technique of point splitting~\cite{Christensen}, the stress tensor operator for a free scalar field can be written in terms of the differential operator~\cite{PH01}
\bea
  \mathcal{T}_{ab}(x,x') &\equiv& \frac{1}{2}(1-2\xi)(g_a^{~a'}\nabla_{a'}\nabla_{b} + g_b^{~b'}\nabla_a\nabla_{b'}) \nn
                                & & + (2\xi-\frac{1}{2})g_{ab}g^{cd'}\nabla_c\nabla_{d'} - \xi (\nabla_a\nabla_b + g_a^{~a'}g_b^{~b'}(\nabla_{a'}\nabla_{b'}) \nn
                                & & + \xi g_{ab} (\nabla_c\nabla^c + \nabla_{c'}\nabla^{c'}) + \xi(R_{ab} - \frac{1}{2}g_{ab}R) - \frac{1}{2}m^2g_{ab}\; \label{t-point-split}
\eea
acting on two copies of the scalar field,
\bea
  \hat{T}_{ab}(x) = \lim_{x' \rightarrow x}\mathcal{T}_{ab}(x,x') [\hat{\phi}(x) \hat{\phi}(x')] \;. \label{xi-def}
\eea
Here, primes on indices indicate tensor indices which transform at the point $x'$, and $g_a^{~a'}$ is the bivector which parallel transports from $x'$ to $x$.  Substituting $\zeta$ for $\phi$ and defining $\xi_{ab}$ as
\bea
  \xi_{ab}(x) = \lim_{x' \rightarrow x} \mathcal{T}_{ab}(x,x') \, [\zeta(x) \zeta(x') - G^+(x,x')] \;, \label{xi-zeta}
\eea
at each order the moments of $\xi_{ab}$ are given by
\bea
  \lag \xi_{a_1b_1}(x_1)...\xi_{a_nb_n}(x_n) \rag_{s} &=& \lag [\lim_{x_1' \rightarrow x_1}\mathcal{T}_{a_1b_1}(x_1,x_1')[\zeta(x_1)\zeta(x_1')] - \lag \hat{T}_{a_1b_1}(x_1) \rag]... \nn
               & & \times [\lim_{x_n' \rightarrow x_n}\mathcal{T}_{a_nb_n}(x_n,x_n')[\zeta(x_n)\zeta(x_n')] - \lag \hat{T}_{a_nb_n}(x_n) \rag]\rag_s \nn
               &=& \lag \{ [\lim_{x_1' \rightarrow x_1}\mathcal{T}_{a_1b_1}(x_1,x_1')[\hat{\phi}(x_1)\hat{\phi}(x_1')] - \lag \hat{T}_{a_1b_1}(x_1) \rag],..., \nn
               & & \times [\lim_{x_n' \rightarrow x_n}\mathcal{T}_{a_nb_n}(x_n,x_n')[\hat{\phi}(x_n)\hat{\phi}(x_n')] - \lag \hat{T}_{a_nb_n}(x_n) \rag] \}\rag \nn
               &=& \lag \{ \hat{t}_{a_1b_1}(x_1),...,\hat{t}_{a_nb_n}(x_n) \} \rag
\eea
satisfying our conjecture.

To compute explicit realizations of the stochastic fluctuations we make use of a Karhunen-Loeve transform~\cite{KLT}, in which the stochastic field $\zeta(x)$ is expanded as
\bea
  \zeta(x) = \sum_i Z_i r^{(i)}(x) \;, \label{klt}
\eea
where each $Z_i$ is an independent Gaussian random variable and each $r^{(i)}$ is a known (non-random) function of $x$.  The functions $r^{(i)}$ satisfy the eigenfunction equation
\bea 
  \int d^4y \, \sqrt{-g(y)} G^+(x,y) r^{(i)}(y) = \lambda_i  r^{(i)}(x) \;, \label{klt-eigenf} 
\eea
and the random coefficients $Z_i$ are determined by
\bes
\bea
  \langle Z_i \rangle_s &=& 0 \\
  \langle Z_i Z_j \rangle_s &=& \delta_{ij} \lambda_i \;. 
\eea \label{z-moments}
\ees

Because of the singular nature of the Wightman function in the coincident limit, it can be shown that no expansion of the form in Eq.~\eqref{klt} exists; the eigenvalues will be infinite and thus $\zeta(x)$ will be infinite in amplitude and discontinuous at every point.  However, this is a manifestation of the well-known pathologies present in the UV behavior of quantum field theory.  For the Wightman function (and other objects quadratic in the field operator) the normal way in which this pathology is dealt with is through renormalization; unfortunately, renormalization fails for the higher correlators due to the presence of non-renormalizable, state dependent divergences.  Instead, regularization of the correlation functions may be obtained through coarse graining, along the lines of~\cite{FFR10}, by integrating the Wightman function $G^+(x,x') \equiv \lag \hat{\phi}(x) \hat{\phi}(x') \rag$ against two copies of a smearing function,
\bea
   \widetilde{G}(x,x') &\equiv& \int d^4x'' d^4x''' \sqrt{-g(x'')}\sqrt{-g(x''')} G^+(x'',x''')W(x''-x)W(x'''-x') \;. \label{smearG}
\eea

Approximate expressions for the coarse-grained $\zeta(x)$ are obtained numerically by discretizing the integral in Eq.~\eqref{klt-eigenf} to a lattice of spacetime points $x_i$,
\bea 
  \sum_j \, \widetilde{G}(x_i,x_j) r^{(k)}(x_j) = \lambda_k  r^{(k)}(x_i) \;. \label{klt-eigenv} 
\eea
Approximate eigenfunctions are recovered from the eigenvectors by
\bea
  r^{(k)}(x) \approx \lambda_k^{-1} \sum_j \, \widetilde{G}(x,x_j) r^{(k)}(x_j) \;,
\eea
giving
\bea
  \zeta(x) \approx \sum_{j,k} Z_k \lambda_k^{-1} \widetilde{G}(x,x_j) r^{(k)}(x_j) \;. \label{zeta-klt}
\eea

We're now in a position to consider the probability distribution of composite functions of $\zeta(x)$.  First, consider the random variable $\chi \equiv \zeta(x_0)^2$ corresponding to the square of the smeared field operator at the point $x_0$.  The characteristic function for $\chi$ is given by
\bea
  \lag e^{i t \chi} \rag_s &=& \lag e^{i t \zeta(x_0)^2} \rag_s \nn
      &=&  1+\sum_{n=1}^{\infty} \frac{(2n-1)!!}{n!}  [i t \widetilde{G}(x_0,x_0)]^n \nn
      &=&  [1-i t \widetilde{G}(x_0,x_0)]^{-\frac{1}{2}} \;. \label{phi2-char}
\eea
This is exactly the characteristic function of a Gamma distributed random variable,
\bea
  P(\chi ) =  \theta(\chi) \frac{\beta^\alpha}{\Gamma(\alpha)} \chi^{\alpha-1} e^{-\beta \chi } \label{phi2-pdf}
\eea
with parameters
\bea
  \alpha=1/2, & & \beta=[2\widetilde{G}(x_0,x_0)]^{-1} \;. \label{phi2-pdf-param}
\eea

For fluctuations of the stress energy tensor, the situation is somewhat different.  The characteristic function for $\xi_{ab}$ may be written as
\bea
 & & \!\!\!\!\!\!\!\!\!\!\!\! \left\lag \exp \left\{ i \int \! d^4x \; \sqrt{-g(x)} \Delta h^{ab}(x) \lim_{x' \rightarrow x} \mathcal{T}_{ab}(x,x') [ \zeta(x) \zeta(x') - \widetilde{G}(x,x') ] \right\} \right\rag_s \nn
 & & = \left\lag \exp \left\{ i \int \! d^4x \; \sqrt{-g(x)} \Delta h^{ab}(x) \left[ \sum_{i,i'} Z_i Z_{i'} T_{ab}^{ii'}(x) - \widetilde{T}_{ab}(x) \right] \right\} \right\rag_s  \nn
 & & = 1 - \frac{1}{2} \int \! d^4x_1 d^4x_2 \; \sqrt{-g(x_1)}\sqrt{-g(x_2)} \Delta h^{ab}(x_1) \Delta h^{cd}(x_2) \sum_{i,j} \lambda_i \lambda_j T_{ab}^{ij}(x_1) T_{cd}^{ij}(x_2) \nn
 & & \;\;\;\; + O(\Delta h^3) \;, \label{xi-char-T}
\eea
where we have used
\bea
  T_{ab}^{ij}(x) &\equiv& \lim_{x' \rightarrow x} \mathcal{T}_{ab}(x,x') [r^{(i)}(x) r^{(j)}(x')] \;, \nn
  \widetilde{T}_{ab}(x) &\equiv& \lim_{x' \rightarrow x} \mathcal{T}_{ab}(x,x')[\widetilde{G}(x,x')] \;.
\eea
Finding a closed form expression, as in Eq.~\eqref{phi2-char}, is difficult for the general case as the coefficients $T_{ab}^{ij}$ will depend on the geometry of the background spacetime.  

However, if we limit consideration to measurements of the vacuum energy density at a single point for a massless scalar field in Minkowski spacetime such a solution exists. Using
\bea
  \lim_{x' \rightarrow x} \mathcal{T}_{0}^{~0} [ \zeta(x) \zeta(x') ] = \frac{1}{2}[\dot{\zeta}(x)^2 + \zeta_{,i}(x) \zeta^{,i}(x)] \;,
\eea
the characteristic function for $\xi_{0}^{~0}(x_0)$ may be written as
\bea
  \lag e^{i t \xi_{0}^{~0}(x_0) } \rag_s = e^{-i t \widetilde{T}_{0}^{~0}(x_0)} \lag \exp \{ i t  \frac{1}{2}[\dot{\zeta}(x)^2+ \zeta_{,i}(x_0) \zeta^{,i}(x_0)] \} \rag_s \;.
\eea
It is easily shown that this expression factorizes by noting that the off diagonal terms of $\widetilde{T}_{ab}(x_0)$ vanish, giving
\bea
  \lag e^{i t \xi_{0}^{~0}(x_0) } \rag_s &=& e^{-i t \widetilde{T}_{0}^{~0}(x_0)} \lag e^{ \frac{i}{2} t \dot{\zeta}(x)^2) } \rag_s \prod_i \lag e^{ \frac{i}{2} t \zeta_{,i}(x_0)^2 } \rag_s \nn
  &=& e^{-i t \widetilde{T}_{0}^{~0}(x_0)} (1 - i t \widetilde{T}_{0}^{~0}(x_0))^{-\frac{1}{2}} (1 - \frac{i}{3} t \widetilde{T}_{0}^{~0}(x_0))^{-\frac{3}{2}} \;. \label{xi-char-Mink}
\eea
The probability density function satisfying this result is
\bea
  P[\xi_0^{~0}(x_0)] &=& \theta(c_1) c_1 c_2 e^{-2c_1} [I_0(c_1) - I_1(c_1)]  \label{xi-pdf}
\eea
where $I_n(x)$ is the modified Bessel function of the first kind and we have used
\bea
   c_1 = 1 + \frac{\xi_0^{~0}(x_0)}{\widetilde{T}_{0}^{~0}(x_0)} \;, & & c_2 = \frac{3^{3/2}}{ \widetilde{T}_{0}^{~0}(x_0)} \;. \label{xi-pdf-param}
\eea

Similarly to the Gamma distribution, this distribution exhibits a sharp cutoff with a minimum located at $\xi_0^{~0}(x_0) = -\widetilde{T}_{0}^{~0}(x_0)$ and a long positive tail, with negative values for the energy density occurring approximately 62\% of the time balanced by rarer but correspondingly larger positive values.  This behavior is markedly different from that predicted by Gaussian fluctuations.  

The effect of $\widetilde{G}(x_0,x_0)]$ and $\widetilde{T}_{0}^{~0}(x_0)$ present in Eqs.~\eqref{phi2-pdf-param} and~\eqref{xi-pdf-param} are to control the amplitude of the fluctuations, with more sharply peaked smearing functions leading to higher amplitudes.  However, if we reparameterize $\chi$ and $\xi_{0}^{~0}$ in terms of the dimensionless quantities
\bea
  \bar{\chi} \equiv \frac{\chi}{\widetilde{G}(x_0,x_0)} \;, & & \bar{\xi}_{0}^{~0} \equiv \frac{\xi_{0}^{~0}}{\widetilde{T}_{0}^{~0}(x_0,x_0)} \;,
\eea
then the corresponding probability distributions are independent of the choice of smearing function,
\bes
\bea
 & P(\bar{\chi} )  = \frac{\bar{\beta}^{\bar{\alpha}} }{ \Gamma(\bar{\alpha}) } \bar{\chi}^{\bar{\alpha}-1} e^{-\bar{\beta} \bar{\chi} } \\
 & \bar{\alpha}=1/2 \;, \;\;\;\;  \bar{\beta}=1/2 \;,
\eea
\ees
and
\bes
\bea
 & P[\bar{\xi}_0^{~0}(x_0)] = \theta(\bar{c}_1) \bar{c}_1 \bar{c}_2 e^{-2\bar{c}_1} [I_0(\bar{c}_1) - I_1(\bar{c}_1)]  \\
  & \bar{c}_1 = 1 + \xi_0^{~0}(x_0) \;, \;\;\;\; \bar{c}_2 = 3^{3/2} \;.
\eea
\ees
In this way, we feel that the probability distributions presented here point to something fundamental about the nature of quantum fluctuations of the stress energy tensor, and are not artifacts of the coarse graining procedure.

Although qualitatively these results are similar to the results found in~\cite{FFR12}, we note disagreement in the details of the distributions.  For fluctuations of $\hat{\phi}^2$, they found a Gamma distribution with $\alpha=1/72$ (compared to our value of $\alpha=1/2$), giving a substantially higher probability for finding negative values - approximately 84\%, compared with 68\% for our method.  For stress energy fluctuations, the disagreement is more pronounced, in that they found a distribution whose moments grow like $C D^n (3n-4)!$ violating the Stieltjes uniqueness criterion for a probability distribution on a half-line; in constrast, numerical evidence suggests the moments of Eq.~\eqref{xi-char-Mink} grow as $C D^n n!$ with $D \approx 1$, indicating that Eq.~\eqref{xi-pdf} is the unique probability density function for $\xi_0^{~0}(x_0)$.

While the discussion so far has focused on the probability of measuring given values at a single point, Equations~\eqref{xi-def} and~\eqref{zeta-klt} provide a roadmap for numerically computing realizations of $\xi_{ab}$ across the entire spacetime, such as would be required for solving the stochastic backreaction problem implied by Eq.~\eqref{e-l}.  Such solutions would be of considerable interest to the study of density fluctuations in inflationary cosmology, where the backreaction of the matter fields could potentially yield a non-Gaussian signature detectable in the cosmic microwave background.  A second application arises in black hole physics, where fluctuations of Hawking radiation provide insight into black hole evaporation~\cite{HuRou07}.  Lastly, stress energy fluctuations are thought to play an important role in assessing the validity of the semiclassical approximation~\cite{kuo-ford, And-Mol-Mot-1}.  In each case, the key is in the details of the small scale structure of spacetime, where quantum effects become dominant.

Acknowlegdements: The author would like to thank Paul Anderson, Bei-Lok Hu, and Hing Tong Cho for helpful conversations.  This work was supported in part by the National Science Council of the Republic of China under the grants NSC 99-2112-M-032-003-MY3, NSC 101-2811-M-032-005, and by the National Center for Theoretical Sciences (NCTS).

\end{document}